\documentclass[12pt,onecolumn, draftcls]{IEEEtran}

\usepackage{amsmath}
\usepackage{amssymb}
\usepackage[dvips]{graphicx}
\usepackage{setspace}
\usepackage{epsfig}

\newcommand{\tSNR}{\text{SNR}}
\newcommand{\I}{\bar{I}}

\newcommand{\figsize}{0.55}

\newtheorem{theo}{Theorem}

\begin{document}

\title{Cognitive Radio Transmission under QoS Constraints and Interference Limitations}
\author{
\vspace{.2cm}
\authorblockN{Sami Akin and
Mustafa Cenk Gursoy}
\thanks{The authors are with the Department of Electrical
Engineering, University of Nebraska-Lincoln, Lincoln, NE, 68588
(e-mails: samiakin@huskers.unl.edu, gursoy@engr.unl.edu).}
\thanks{This work was supported by the National Science Foundation under
Grants CNS -- 0834753 and CCF--0917265.}}
\date{}

\maketitle

\vspace{-.5cm}
\begin{abstract}
In this paper, the performance of cognitive transmission under quality of service (QoS)
constraints and interference limitations is studied. Cognitive secondary users are assumed to initially perform sensing over multiple frequency bands (or equivalently channels) to detect the activities of primary users. Subsequently, they perform transmission in a single channel at variable power and rates depending on the channel sensing decisions and the fading environment. A state transition model is constructed to
model this cognitive operation. Statistical limitations on the buffer lengths are imposed to take into account the QoS constraints of the cognitive secondary users. Under such QoS constraints and limitations on the interference caused to the primary users, the maximum throughput is identified by finding the effective capacity of the cognitive radio channel. Optimal power allocation strategies are obtained and the optimal channel selection criterion is identified. The intricate interplay between effective capacity, interference and QoS constraints, channel sensing parameters and reliability, fading, and the number of available frequency bands is investigated through numerical results.


\end{abstract}

\begin{spacing}{1.65}

\section{Introduction}
Recent years have witnessed much interest in cognitive radio systems
due to their promise as a technology that enables systems to utilize
the available spectrum much more effectively. This interest has
resulted in a spur of research activity in the area. In \cite{Vahid}, Asghari and Aissa, under constraints on the average interference caused at the
licensed user over Rayleigh fading channels,
studied two adaptation policies at the secondary user's transmitter
in a cognitive radio system one of which is variable power and the
other is variable rate and power. They maximized the achievable
rates under the above constraints and the bit error rate (BER)
requirement in MQAM modulation. The authors in \cite{Leyla'lar} derived
the fading channel capacity of a secondary user subject to both
average and peak received-power constraints at the primary receiver.
In addition, they obtained optimum power allocation schemes for
three different capacity notions, namely, ergodic, outage, and
minimum-rate. Ghasemi \textit{et al.} in \cite{Amir} studied the performance of
spectrum-sensing radios under channel fading. They
showed that due to uncertainty resulting from fading, local signal
processing alone may not be adequate to meet the performance
requirements. Therefore, to remedy this uncertainty they also
focused on the cooperation among secondary users and the tradeoff
between local processing and cooperation in order to maximize the
spectrum utilization. Furthermore, the authors in \cite{liang}
focused on the problem of designing the sensing duration to maximize
the achievable throughput for the secondary network under the
constraint that the primary users are sufficiently protected. They
formulated the sensing-throughput tradeoff problem mathematically,
and use energy detection sensing scheme to prove that the formulated
problem indeed has one optimal sensing time which yields the highest
throughput for the secondary network. Moreover, Quan \textit{et al.}
in \cite{poor2} introduced a novel wideband spectrum sensing
technique, called multiband joint detection, which jointly detects
the signal energy levels over multiple frequency bands rather than
considering one band at a time.

In many wireless systems, it is very important to provide reliable
communications while sustaining a certain level of
quality-of-service (QoS) under time-varying channel conditions. For
instance, in wireless multimedia transmissions, stringent delay QoS
requirements need to be satisfied in order to provide acceptable
performance levels. In cognitive radio systems, challenges in
providing QoS assurances increase due to the fact that secondary
users should operate under constraints on the interference levels
that they produce to primary users. For the secondary users, these
interference constraints lead to variations in transmit power levels
and channel accesses. For instance, intermittent access to the
channels due to the activity of primary users make it difficult for
the secondary users to satisfy their own QoS limitations.

These considerations have led to studies that investigate the
cognitive radio performance under QoS constraints. Musavian and
Aissa in \cite{Leyla'm} considered variable-rate, variable-power MQAM
modulation employed under delay QoS constraints over
spectrum-sharing channels. As a performance metric, they used the effective capacity to characterize the maximum throughput under
QoS constraints. They assumed two users sharing the spectrum with
one of them having a primary access to the band. The other,
known as secondary user, is constrained by interference limitations
imposed by the primary user. Considering two modulation schemes,
continuous MQAM and discrete MQAM with restricted constellations,
they obtained the effective capacity of the secondary user's link,
and derived the optimum power allocation scheme that maximizes the
effective capacity in each case. Additionally, in \cite{Leyla}, they
proposed a QoS constrained power and rate allocation scheme for
spectrum sharing systems in which the
secondary users are allowed to use the spectrum under an
interference constraint by which a minimum-rate of transmission is
guaranteed to the primary user for a certain percentage of time.
Moreover, applying an average interference power constraint which is
required to be fulfilled by the secondary user, they obtained the
maximum arrival-rate supported by a Rayleigh block-fading channel
subject to satisfying a given statistical delay QoS constraint. We note that in these studies on the performance under QoS limitations, channel sensing is not incorporated into the system model. As a result, adaptation of the cognitive transmission according to the presence or absence of the primary users is not considered.

In \cite{Onceki Paper}, where we also concentrated on cognitive
transmission under QoS constraint, we assumed that the
secondary transmitter sends the data at two different fixed rates
and power levels, depending on the activity of the primary users,
which is determined by channel sensing performed by the secondary
users. We constructed a state transition model with eight states to
model this cognitive transmission channel, and determined the effective capacity. On the other hand, we assumed in \cite{Onceki Paper} that channel sensing is done only in one channel, and did not impose explicit interference constraints.

In this paper, we study the effective capacity of cognitive radio
channels where the cognitive radio detects the activity of primary
users in a multiband environment and then performs the data
transmission in one of the transmission channels. Both the secondary
receiver and the secondary transmitter know the fading coefficients
of their own channel, and of the channel between the secondary
transmitter and the primary receiver. The cognitive radio has two
power allocation policies depending on the activities of the primary
users and the sensing decisions. More specifically, the
contributions of this paper are the following:
\begin{enumerate}
  \item We consider a scenario in which the cognitive system employs
  multi-channel sensing and uses one channel for
  data transmission thereby decreasing the probability of
  interference to the primary users.
  \item We identify a state-transition model for cognitive radio
  transmission in which we compare the transmission rates with
  instantaneous channel capacities, and also incorporate the results
  of channel sensing.
  \item We determine the effective capacity of the cognitive channel
  under limitations on the average interference power experienced by
  the primary receiver.

  \item We identify the optimal criterion to select the transmission channel out of the available channels and obtain the optimal power adaptation policies that maximize the effective capacity.

  \item We analyze the interactions between the effective capacity,
  QoS constraints, channel sensing duration, channel detection
  threshold, detection and false alarm probabilities through
  numerical techniques.
\end{enumerate}

The organization of the rest of the paper is as follows: In Section
\ref{system}, we discuss the channel model and analyze
multi-channel sensing. We describe the channel state transition
model in Section \ref{subsec:state} under the assumption that the
secondary users have perfect CSI and send the data at rates equal
to the instantaneous channel capacity values. In Section \ref{sec:consraint},
we analyze the received interference power at the
primary receiver and apply this as a power constraint on the
secondary users. In Section \ref{effective capacity}, we
define the effective capacity and find the optimal power
distribution and show the criterion to choose the best channel.
Numerical results are shown in Section \ref{NumericalResults}, and
conclusions are provided in Section \ref{Conlusion}.

\section{Cognitive Channel Model and Channel Sensing} \label{system}
In this paper, we consider a cognitive radio system in which secondary
users sense $M$ channels and choose one channel for data
transmission. We assume that channel sensing and data transmission
are conducted in frames of duration $T$ seconds. In each frame, $N$
seconds is allocated for channel sensing while data transmission
occurs in the remaining $T-N$ seconds. Transmission power and rate
levels depend on the primary users' activities. If all of the
channels are detected as busy, transmitter selects one channel with
a certain criterion
, and sets the transmission power and rate to $P_{k,1}(i)$ and
$r_{k,1}(i)$, respectively, where $k\in\{1,2,\dots,M\}$ is the index
of the selected channel and $i=1,2,\dots$ denotes the time index.
Note that if $P_{k,1}(i)=0$, transmitter stops sending information
when it detects primary users in all channels. If at least one
channel is sensed to be idle, data transmission is performed with
power $P_{k,2}(i)$ and at rate $r_{k,2}(i)$. If multiple channels
are detected as idle, then one idle channel is selected again
considering a certain criterion.

The discrete-time channel input-output relation between the
secondary transmitter and receiver in the $i^{\text{th}}$ symbol
duration in the $k^{\text{th}}$ channel is given by
\begin{align}\label{input-out1}
&y_{k}(i)=h_{k}(i)x_{k}(i)+n_{k}(i)\quad i=1,2,\dots,
\end{align}
if the primary users are absent. On the other hand, if primary users
are present in the channel, we have
\begin{align}\label{input-out2}
&y_{k}(i)=h_{k}(i)x_{k}(i)+s_{k,p}(i)+n_{k}(i)\quad i=1,2,\dots,
\end{align}
where $x_{k}(i)$ and $y_{k}(i)$ denote the complex-valued channel
input and output, respectively. In (\ref{input-out1}) and
(\ref{input-out2}), $h_{k}(i)$ is the channel fading coefficient between the
cognitive transmitter and the receiver. We assume that $h_{k}(i)$ has a finite variance, i.e.,
$\sigma_{h_{k}}^{2} < \infty$, but otherwise has an arbitrary distribution. We define $z_{k}(i)=|h_{k}(i)|^{2}$. We
consider a block-fading channel model and assume that the fading
coefficients stay constant for a block of duration $T$ seconds and
change from one block to another independently in each channel. In
(\ref{input-out2}), $s_{k,p}(i)$ represents the active primary
user's faded signal arriving at the secondary receiver in the
$k^{th}$ channel, and has a variance $\sigma_{s_{k,p}}^{2}(i)$.
$n_{k}(i)$ models the additive thermal noise at the receiver, and is
a zero-mean, circularly symmetric, complex Gaussian random variable
with variance $\mathbb{E}\{|n_{k}(i)|^{2}\}=\sigma_{n_{k}}^{2}$ for
all $i$. We assume that the bandwidth of the $k^{th}$
channel is $\textit{B}_{k}$.

\begin{figure}
\begin{center}
\includegraphics[width = \figsize\textwidth]{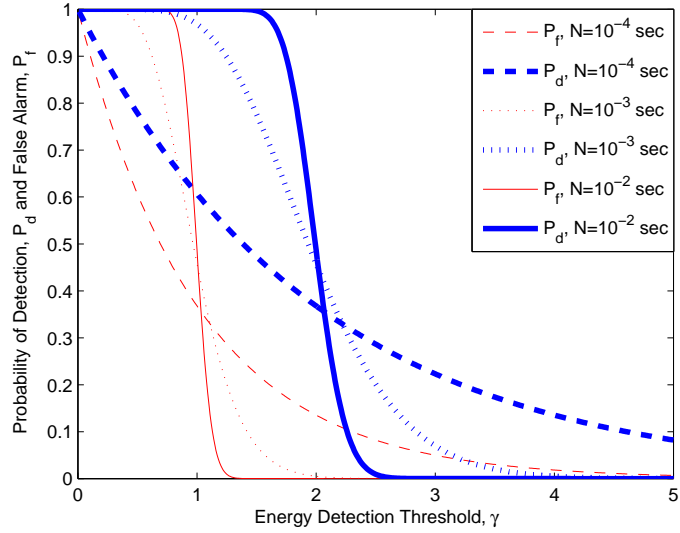}
\caption{Probability of Detection
$P_{d}$ and False Alarm $P_{f}$ vs. Energy Detection Threshold} \label{fig:fig1}
\end{center}
\end{figure}

In the absence of detailed information on primary users'
transmission policies, energy-based detection methods are favorable
for channel sensing. Knowing that wideband channels exhibit
frequency selective features, we can divide the band into channels
and estimate each received signal through its discrete Fourier
transform (DFT) \cite{poor2}. The channel sensing can be formulated
as a hypothesis testing problem between the noise $n_{k}(i)$ and the
signal $s_{k,p}(i)$ in noise. Noting that there are $NB_{k}$ complex
symbols in a duration of $N$ seconds in each channel with bandwidth
$B_{k}$, the hypothesis test in channel $k$ can mathematically be
expressed as follows:
\begin{align}\label{hypothesis}
\begin{split}
&\mathcal{H}_{k,0}\quad : \quad y_{k}(i)=n_{k}(i), \quad
i=1,\dots,NB_{k}\\ &\mathcal{H}_{k,1}\quad : \quad
y_{k}(i)=s_{k,p}(i)+n_{k}(i), \quad i=1,\dots,NB_{k}.
\end{split}
\end{align}
For the above detection problem, the optimal Neyman-Pearson detector
is given by \cite{Poor-book}
\begin{equation}\label{Neyman-Pearson}
Y_{k}=\frac{1}{NB_{k}}\sum_{i=1}^{NB_{k}}|y_{k}(i)|^{2}\gtrless^{\mathcal{H}_{k,1}}_{\mathcal{H}_{k,0}}\gamma_{k}.
\end{equation}
We assume that $s_{k,p}(i)$ has a circularly symmetric complex
Gaussian distribution with zero-mean and variance
$\sigma_{s_{k,p}}^{2}$.
Assuming further that $\{s_{k,p}(i)\}$ are i.i.d., we can
immediately conclude that the test statistic $Y_{k}$ is chi-square
distributed with $2NB_{k}$ degrees of freedom. In this case, the
probabilities of false alarm and detection can be established as
follows:
\begin{align}\label{false alarm}
&P_{k,f}=Pr(Y_{k}>\gamma_{k}|\mathcal{H}_{k,0})=1-P\left(\frac{NB_{k}
\gamma_{k}}{\sigma_{n_{k}}^{2}},NB_{k}\right)\\
&P_{k,d}=Pr(Y_{k}>\gamma_{k}|\mathcal{H}_{k,1})=1-P\left(\frac{NB_{k}
\gamma_{k}}{\sigma_{n_{k}}^{2}+\sigma_{s_{k,p}}^{2}},NB_{k}\right)
\label{eq:probdetect}
\end{align}
where $P(x,a)$ denotes the regularized lower gamma function and is
defined as $P(x,a) = \frac{\gamma(x,a)}{\Gamma(a)}$ where
$\gamma(x,a)$ is the lower incomplete gamma function and $\Gamma(a)$
is the Gamma function. In Figure \ref{fig:fig1}, the probability of
detection, $P_{d}$, and the probability of false alarm, $P_{f}$, are plotted as a
function of the energy detection threshold, $\gamma$, for different
values of channel detection duration. Note that the bandwidth
is $B=10$kHz and the block duration is $T=0.1$s. We can see that when the
detection threshold is low, $P_{d}$ and $P_{f}$ tend to be 1, which
means that the secondary user, always assuming the existence of an active primary user, transmits with power $P_{1}(i)$ and
rate $r_{1}(i)$. On the other hand, when the detection threshold is
high, $P_{d}$ and $P_{f}$ are close to zero, which means that the
secondary user, being unable to detect the activity of the primary users, always transmits with power $P_{2}(i)$ and rate
$r_{2}(i)$, possibly causing significant interference. The main purpose is to keep $P_{d}$ as close to 1 as
possible and $P_{f}$ as close to 0 as possible. Therefore, we have
to keep the detection threshold in a reasonable interval. Note that
the duration of detection is also important since increasing the
number of channel samples used for sensing improves the quality of
channel detection.

In the hypothesis testing problem in (\ref{hypothesis}), another approach is to
consider $Y_{k}$ as Gaussian distributed, which is accurate if
$NB_{k}$ is large \cite{liang}. In this case, the detection and
false alarm probabilities can be expressed in terms of Gaussian
$Q$-functions. We would like to note the rest of the analysis in the paper does not depend on the specific expressions of the false alarm and detection probabilities. However, numerical results are obtained using (\ref{false alarm}) and (\ref{eq:probdetect}).

\section{State Transition Model} \label{subsec:state}
In this paper, we assume that both the secondary receiver and
transmitter have perfect channel side information (CSI), and hence
perfectly know the realizations of the fading coefficients
$\{h_{k}(i)\}$. We further assume that the wideband channel is
divided into channels, each with bandwidth that is equal to the
coherence bandwidth $B_{c}$. Therefore, we henceforth have
$B_{k}=B_{c}$. With this assumption, we can suppose that independent
flat fading is experienced in each channel. In order to further
simplify the setting, we consider a symmetric model in which fading
coefficients are identically distributed in different channels.
Moreover, we assume that the
background noise and primary users' signals are also identically
distributed in different channels and hence their variances
$\sigma_{n}^{2}$ and $\sigma_{s_{p}}^{2}$ do not depend on $k$, and
the prior probabilities of each channel being occupied by the
primary users are the same and equal to $\rho$. In channel sensing,
the same energy threshold, $\gamma$, is applied in each channel.
Finally, in this symmetric model, the transmission power and rate
policies when the channels are idle or busy are the same for each
channel. Due to the consideration of a symmetric model, we in the
subsequent analysis drop the subscript $k$ in the expressions for
the sake of brevity.

First, note that we have the following four possible scenarios
considering the correct detections and errors in channel sensing:
\begin{description}
  \item{\emph{Scenario 1:}} All channels are detected as busy, and channel used for transmission is actually busy.
  \item{\emph{Scenario 2:}} All channels are detected as busy, and channel used for transmission is actually idle.
  \item{\emph{Scenario 3:}} At least one channel is detected as idle, and channel used for transmission is actually busy.
  \item{\emph{Scenario 4:}} At least one channel is detected as idle, and channel used for transmission is actually idle.
\end{description}
In each scenario, we have one state, namely either ON or OFF,
depending on whether or not the instantaneous transmission rate
exceeds the instantaneous channel capacity. Considering the
interference $s_{p}(i)$ caused by the primary users as additional
Gaussian noise, we can express the instantaneous channel capacities
in the above four scenarios as follows:
\begin{description}\label{channel capacity}
\item{\emph{Scenario 1:}} $C_{1}(i)=B_{c}\log_{2}(1+\tSNR_{1}(i))$.
\item{\emph{Scenario 2:}} $C_{2}(i)=B_{c}\log_{2}(1+\tSNR_{2}(i))$.
\item{\emph{Scenario 3:}} $C_{3}(i)=B_{c}\log_{2}(1+\tSNR_{3}(i))$.
\item{\emph{Scenario 4:}} $C_{4}(i)=B_{c}\log_{2}(1+\tSNR_{4}(i))$.
\end{description}
Above, we have defined
\begin{equation}
\tSNR_{1}(i)=\frac{P_{1}(i)z(i)}{B_{c}\left(\sigma_{n}^{2}+\sigma_{s_{p}}^{2}\right)},
\tSNR_{2}(i)=\frac{P_{1}(i)z(i)}{B_{c}\sigma_{n}^{2}},
\tSNR_{3}(i)=\frac{P_{2}(i)z(i)}{B_{c}\left(\sigma_{n}^{2}+\sigma_{s_{p}}^{2}\right)},
\tSNR_{4}(i)=\frac{P_{2}(i)z(i)}{B_{c}\sigma_{n}^{2}}.
\end{equation}
Note that $z(i)=|h(i)|^{2}$ denotes the fading power. In scenarios 1 and 2, the
secondary transmitter detects all channels as busy and transmits
the information at rate
\begin{equation}\label{rate1}
r_{1}(i)=B_{c}\log_{2}\left(1+\tSNR_{1}(i)\right).
\end{equation}
On the other hand, in scenarios 3 and 4, at least one channel is
sensed as idle and the transmission rate is
\begin{equation}\label{rate2}
r_{2}(i)=B_{c}\log_{2}\left(1+\tSNR_{4}(i)\right),
\end{equation}
since the transmitter, assuming the channel as idle, sets the power level to $P_2(i)$ and expects that no interference from the primary transmissions will be experienced at the secondary receiver (as seen by the absence of $\sigma_{s_{p}}^{2}$ in the denominator of $\tSNR_{4}$).

In scenarios 1 and 2, transmission rate is less than or equal to the
instantaneous channel capacity. Hence, reliable transmission at rate
$r_{1}(i)$ is attained and channel is in the ON state. Similarly,
the channel is in the ON state in scenario 4 in which the
transmission rate is $r_{2}(i)$. On the other hand, in scenario 3,
transmission rate exceeds the instantaneous channel capacity (i.e., $r_2(i) > C_3(i)$) due to
miss-detection. In this case, reliable communication cannot be
established, and the channel is assumed to be in the OFF state. Note
that the effective transmission rate in this state is zero, and
therefore information needs to be retransmitted. We assume that this is accomplished through a simple ARQ mechanism.

\begin{figure}
\begin{center}
\includegraphics[width = 0.9\textwidth]{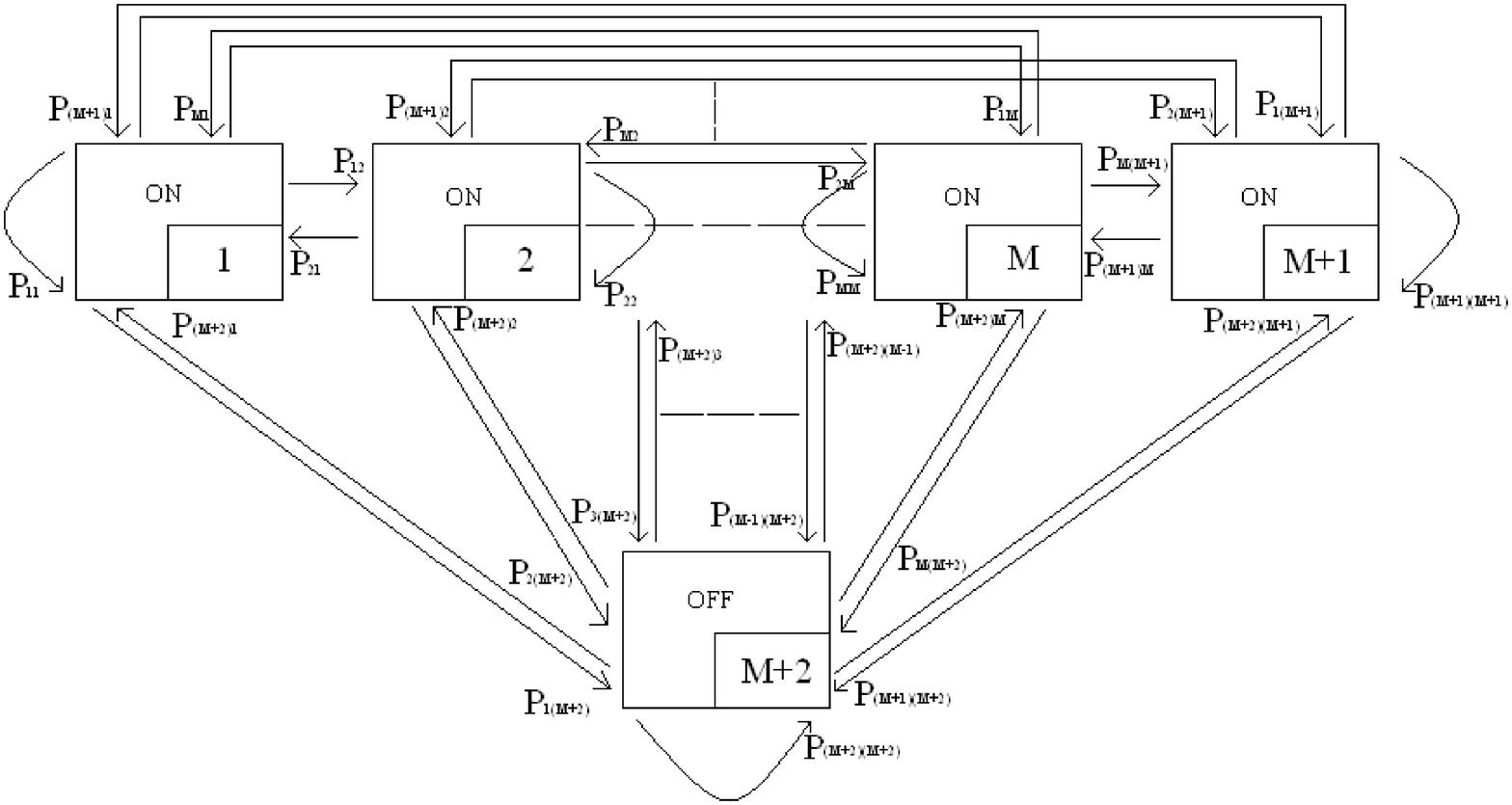}
\caption{State transition model for the cognitive radio channel. The numbered label for each state is given on the
lower-right corner of the box representing the state.}
\label{fig:fig2}
\end{center}
\end{figure}

For this cognitive
transmission model, we initially construct a state transition model.
While the ensuing discussion describes this model, Figure
\ref{fig:fig2} provides a depiction. As seen in Fig. \ref{fig:fig2}, there are $M+1$ ON
states and 1 OFF state. The single OFF state is the one experienced in scenario 3.  The first ON state, which is the top
leftmost state in Fig. \ref{fig:fig2}, is a combined version of the ON states in scenarios 1 and 2 in both of which the transmission rate is
$r_{1}(i)$ and the transmission power is $P_{1}(i)$. Note that
all the channels are detected as busy in this first ON state. The remaining ON
states labeled $2$ through $(M+1)$ can be seen as the expansion of the ON state in scenario 4 in which at least one channel is detected as idle and the channel chosen for transmission is actually idle. More specifically, the $k^{\text{th}}$ ON state for $k = 2,3, \ldots, M+1$ is the ON state in which $k-1$ channels are detected as idle and the channel chosen for transmission is idle. Note that the transmission rate is $r_{2}(i)$
and the transmission power is $P_{2}(i)$ in all ON states labeled $2$ through $(M+1)$.

Next, we characterize the state transition probabilities. State transitions occur every
$T$ seconds. We can easily see that the probability of staying
in the first ON state, in which all channels are detected as busy, is expressed as follows:
\vspace{-.4cm}
\begin{equation} \label{eq:transitionprob}
p_{11}=\alpha^{M}
\end{equation}
where $\alpha=\rho P_{d}+\left(1-\rho\right)P_{f}$ is the
probability that channel is detected as busy, and $P_{d}$ and
$P_{f}$ are the probabilities of detection and false alarm,
respectively as defined in (\ref{eq:probdetect}). Recall that $\rho$ denotes the probability that a channel is busy (i.e., there are active primary users in the channel). It is important to note that the transition probability in (\ref{eq:transitionprob}) is obtained under the assumptions that the primary user activity is independent among the channels and also from one block to another. Indeed, under the assumption of independence over the blocks, the state transition probabilities do not depend on the originating
state and hence we have
\begin{align}
p_{11}=p_{21}=\dots =p_{(M+1)1}=p_{(M+2)1}=\alpha^{M} \triangleq p_1 \label{eq:transitionprob1}
\end{align}
where we have defined $p_1 = p_{i1}$ for all $ i = 1,2,\ldots, M+2$.
Similarly, we can obtain for $k=2,3,\dots,M+1$,
\begin{align}
p_{1k}=p_{2k}=\dots=p_{(M+1)k}=p_{(M+2)k}&= P\left(\substack{\text{$(k-1)$ out of $M$}\\ \text{channels are detected as idle}} \text{ and } \substack{\text{the channel chosen for transmission} \\ \text{is actually idle}} \right)
\\
&=
\underbrace{\left(\begin{array}{c}
M \\ k-1
\end{array}\right) \alpha^{M-k+1} (1-\alpha)^{k-1}}_{\substack{\text{probability that $(k-1)$ out of $M$ channels}\\ \text{are detected as idle}}}\times \underbrace{\frac{(1-\rho)(1-P_f)}{1-\alpha}}_{\substack{\text{probability that the channel chosen for} \\ \text{transmission is actually idle}\\\text{given that it is detected as idle}}} \label{eq:transitionprobformula}
\\
&
=\frac{M!}{(M-k+1)!(k-1)!} \,\alpha^{M-k+1}\left(1-\alpha\right)^{k-2}\left(1-\rho\right)\left(1-P_{f}\right)
\\
&\triangleq p_k \label{eq:transitionprobk}
\end{align}
Now, we can easily observe that the
transition probabilities for the OFF state are
\begin{align}
p_{1(M+2)}=p_{2(M+2)}=\dots
=p_{(M+1)(M+2)}=p_{(M+2)(M+2)}&=1-\sum_{k=1}^{M+1}p_{1k}\\&=\sum_{k=1}^{M}\frac{M!}{(M-k)!k!}\,\alpha^{M-k}\left(1-\alpha\right)^{k-1}\rho(1-P_{d})\nonumber \\
&\triangleq p_{M+2}.\label{eq:transitionprobM+2}
\end{align}
Then, we can easily see
that the $(M+2)\times(M+2)$ state transition probability matrix can
be expressed as
\begin{center}
    $R =
    \left[
        \begin{array}{cccc}
            p_{1,1}     & . & . & p_{1,M+2}  \\
            .&\quad&\quad&.\\
            .&\quad&\quad&.\\
            p_{M+2,1}   & . & . & p_{M+2,M+2}\\
        \end{array}
    \right]
    =
    \left[
        \begin{array}{cccc}
            p_{1}     & . & . & p_{M+2}  \\
            .&\quad&\quad&.\\
            .&\quad&\quad&.\\
            p_{1}     & . & . & p_{M+2}\\
        \end{array}
    \right]$
\end{center}
Note that $R$ has a rank of 1. Note also that in each frame duration
of $T$ seconds, $r_{1}(k)(T-N)$ bits are transmitted and received in
state 1, and $r_{2}(k)(T-N)$ bits are transmitted and received
in states 2 through $M+1$, while the transmitted number of
bits is assumed to be zero in state $M+2$.

\section{Interference Power Constraints} \label{sec:consraint}
In this section, we consider interference power constraints to limit
the transmission powers of the secondary users and provide
protection to primary users. In particular, we assume that the transmission power of the
secondary users is constrained in such a way that the average
interference power on the primary receiver is limited.

Note that interference to the primary users is caused in scenarios 1 and 3. In scenario 1, the channel is busy, and the secondary user, detecting the channel as busy, transmits at power level $P_1$. Consequently, the instantaneous interference power experienced by the primary user is $P_1 z_{sp}$ where $z_{sp} = |h_{sp}(i)|^{2}$ is the magnitude-square of the fading coefficient of the channel between the secondary transmitter and the primary user. Note also that the probability of being in scenario 1 (i.e., the probability of detecting all channels busy and having the chosen transmission channel as actually busy) is $\alpha^{M-1}\rho P_d$, as can be easily seen through an analysis similar to that in (\ref{eq:transitionprobformula}).

In scenario 3, the secondary user, detecting the channel as idle, transmits at power $P_2$ although the channel is actually is busy. In this case, the instantaneous interference power is $P_2 z_{sp}$. Since we consider power adaption, transmission power levels $P_1$ and $P_2$ in general vary with $z_{sp}$ and also with $z$, which is the power of the fading coefficient between the secondary transmitter and secondary receiver in the chosen transmission channel. Hence, in both scenarios, the instantaneous interference power levels depend on both $z_{sp}$ and $z$ whose distributions depend on the criterion with which the transmission channel is chosen and the number of available channels from which the selection is performed. For this reason, it is necessary in scenario 3 to separately consider the individual cases with different number of idle-detected channels. We have $M$ such cases. For instance, in the $k^{th}$ case for $k = 1,2,\ldots, M$, we have $k$ channels detected as idle and the channel chosen out of these $k$ channels is actually busy. The probability of the $k^{th}$ case can be easily found to be
$
\frac{M!}{(M-k)!k!}\, \alpha^{M-k}\left(1-\alpha\right)^{k-1}\rho(1-P_{d}).
$

Following the above discussion, we can now express the average interference constraints as follows:
\begin{align}\label{average interference power}
\underbrace{\alpha^{M-1}\rho P_{d}}_{\substack{\text{probability of}\\\text{scenario 1}}} \,\,
\underbrace{E_{}\left\{P_{1}z_{sp}\right\}}_{\substack{\text{average interference} \\ \text{in scenario 1}}}+\sum_{k=1}^{M}\underbrace{\frac{M!}{(M-k)!k!}\,\alpha^{M-k}\left(1-\alpha\right)^{k-1}\rho(1-P_{d})}_{\text{probability of the $k^{th}$ case of scenario 3}}
\underbrace{E_{k}\left\{P_{2}z_{sp}\right\}}_{\substack{\text{average interference}\\\text{in the $k^{th}$ case}\\\text{of scenario 3}}} \leq I_{avg}
\end{align}
Note from above that $I_{avg}$ is the constraint on the interference averaged over the distributions of $z$ and $z_{sp}$ (through the expectations), and also averaged over the probabilities of different scenarios and cases. It is important to note that the term $E_k\left\{P_{2}z_{sp}\right\}$, as discussed above, depends in general on the number of idle-detected channels, $k$. This dependence is indicated through the subscript $k$.

In a system with more strict requirements on the interference, the following individual interference constraints can be imposed:
\begin{align}
E_{}\left\{P_{1}z_{sp}\right\} &\le I_0 \quad \text{ and } \quad
E_{k}\left\{P_{2}z_{sp}\right\} \le I_k \quad \text{for } k = 1,2,\ldots, M. \label{eq:individualinterference}
\end{align}
If, for instance, $I_0 = I_1 = I_2 = \ldots = I_M$, then interference averaged over fading is limited by the same constraint regardless of which scenario is being realized. As considered in \cite{Leyla}, by appropriately choosing the values of $I_0$ and $I_k$ in (\ref{eq:individualinterference}), we can provide primary users a minimum rate guarantee for a certain percentage of the time in a Rayleigh fading environment through the following outage constraints:
\begin{align}
&Pr\left\{\log_{2}\left(1+\frac{P_{pri}z_{p}(i)}{P_{1}(i)z_{sp}(i)+\sigma_{n_{p}}^{2}B_{c}}\right)\leq
R_{min}\right\}\leq P_{1}^{out}, \label{outage1}\\
&Pr\left\{\log_{2}\left(1+\frac{P_{pri}z_{p}(i)}{P_{2}(i)z_{sp}(i)+\sigma_{n_{p}}^{2}B_{c}}\right)\leq
R_{min}\right\}\leq P_{2,k}^{out}. \quad \text{for } k = 1,2, \ldots, M.\label{outage2}
\end{align}
$P_{1}^{out}$ and $P_{2,k}^{out}$ can be seen as the outage constraints in scenario 1 and in the $k^{th}$ case of scenario 3, respectively. In the above formulations, $R_{min}$ is the required minimum
transmission rate to be provided to the primary users with outage
probabilities $P_{1}^{out}$ and $P_{2,k}^{out}$, and
$z_{p}(i)=|h_{p}(i)|^{2}$ where $h_p$ is the fading coefficient of
the channel between the primary transmitter and primary receiver.  $\sigma_{n_{p}}^{2}$ is the variance of the zero-mean, circularly
symmetric, complex Gaussian thermal noise at the primary receiver. $P_{pri}$ is the transmission power of the primary transmitter. Under the assumption that $z_p$ is an exponential random variable (i.e., we have a Rayleigh fading channel between the primary transmitter and receiver), the outage probability in (\ref{outage1}) can be expressed as follows:
\begin{align}
Pr\left\{\log_{2}\left(1+\frac{P_{pri}z_{p}(i)}{P_{1}(i)z_{sp}(i)+\sigma_{n_{p}}^{2}B_{c}}\right)\leq
R_{min}\right\} &= Pr\left\{z_{p}\leq \frac{2^{R_{min}}-1}{P_{pri}}\left(P_{1}(i)z_{sp}(i)+\sigma_{n_{p}}^{2}B_{c}\right)\right\} \label{eq:outageconstraint1}
\\
&=E\left\{1-e^{-\frac{2^{R_{min}}-1}{P_{pri}}(P_{1}(i)z_{sp}(i)+\sigma_{n_{p}}^{2}B_{c})}\right\}\label{eq:outageconstraint2}
\\
&\le 1-e^{-\frac{2^{R_{min}}-1}{P_{pri}}\left(E\left\{P_{1}(i)z_{sp}(i)\right\}+\sigma_{n_{p}}^{2}B_{c}\right)}\label{eq:outageconstraint3}
\end{align}
where (\ref{eq:outageconstraint2}) is obtained by performing integration with respect to the probability density function (pdf) of $z_p$ in the evaluation of the probability expression in (\ref{eq:outageconstraint1}).   As a result, the expectation in (\ref{eq:outageconstraint2}) is with respect to the remaining random components $P_1$ and $z_{sp}$. Finally, the inequality in (\ref{eq:outageconstraint3}) follows from the concavity of the function $1-e^{-x}$ and Jensen's inequality. From (\ref{eq:outageconstraint3}), we can immediately see that if we impose
\begin{gather}
E_{}\left\{P_{1}z_{sp}\right\}\leq\Phi_{1}=-\frac{\log_{e}\left(1-P_{1}^{out}\right)}{\frac{2^{R_{min}}-1}{P_{pri}}}-\sigma_{n_{p}}^{2}B_{c},
\end{gather}
then the constraint in (\ref{outage1}) will be satisfied. A  similar discussion follows for (\ref{outage2}) as well.

In the subsequent parts of the paper, we assume that an average interference power constraint in the form given in (\ref{average interference power}) is imposed.

\section{Effective Capacity}\label{effective capacity}

In this section, we identify the maximum throughput that the
cognitive radio channel with the aforementioned state-transition
model can sustain under interference power constraints and statistical QoS limitations imposed in the
form of buffer or delay violation probabilities\footnote{Note that interference constraints are imposed to provide a certain level of quality-of-service to the primary users, while buffer or delay constraints are used to statistically guarantee a quality-of-service level to the transmissions of the secondary users. Hence, the formulation in the paper effectively considers service guarantees for both the primary and secondary users. On the other hand, QoS constraints throughout the paper refer to buffer/delay constraints to avoid confusion.}. Wu and Negi in
\cite{Wu} defined the effective capacity as the maximum constant
arrival rate that can be supported by a given channel service
process while also satisfying a statistical QoS requirement
specified by the QoS exponent $\theta$. If we define $Q$ as the
stationary queue length, then $\theta$ is defined as the decay rate
of the tail distribution of the queue length $Q$:
\begin{equation}\label{decayrate}
\lim_{q\rightarrow \infty}\frac{\log P(Q\geq q)}{q}=-\theta.
\end{equation}
Hence, we have the following approximation for the buffer violation
probability for large $q_{max}$: $P(Q\geq q_{max})\approx e^{-\theta
q_{max}}$. Therefore, larger $\theta$ corresponds to more strict QoS
constraints, while the smaller $\theta$ implies looser constraints.
In certain settings, constraints on the queue length can be linked
to limitations on the delay and hence delay-QoS constraints. It is
shown in \cite{Liu} that $P\{D\geq d_{max}\}\leq c\sqrt{P\{Q\geq
q_{max}\}}$ for constant arrival rates, where $D$ denotes the
steady-state delay experienced in the buffer. In the above
formulation, $c$ is a positive constant, $q_{max}=ad_{max}$ and $a$
is the source arrival rate. Therefore, effective capacity provides
the maximum arrival rate when the system is subject to statistical
queue length or delay constraints  in the forms of $P(Q \ge
q_{\max}) \le e^{-\theta q_{max}}$ or $P\{D \ge d_{\max}\} \le c \,
e^{-\theta a \, d_{max}/2}$, respectively. Since the average arrival
rate is equal to the average departure rate when the queue is in
steady-state \cite{ChangZajic}, effective capacity can also be seen
as the maximum throughput in the presence of such constraints.

The effective capacity for a given QoS exponent $\theta$ is given by
\begin{equation}\label{exponent}
-\lim_{t\rightarrow \infty}\frac{1}{\theta
t}\log_{e}E\{e^{-\theta
S(t)}\}=-\frac{\Lambda(-\theta)}{\theta}
\end{equation}
where $S(t)=\sum_{k=1}^{t}r(k)$ is the time-accumulated service
process, and $\{r(k),k=1,2,\dots\}$ is defined as the discrete-time,
stationary and ergodic stochastic service process. Note that
$\Lambda(\theta)$ is the asymptotic log-moment generating function
of $S(t)$, and is given by
\begin{equation}\label{asymptotic log-moment}
\Lambda(\theta)=\lim_{t\rightarrow \infty}\frac{1}{t}\log
E\left[e^{\theta S(t)}\right].
\end{equation}
The service rate according to the model described in Section \ref{subsec:state} is
$r(k) = r_{1}(k)(T-N)$ if the cognitive system is in state 1 at time
$k$. Similarly, the service rate is $r(k) = r_{2}(k)(T-N)$ in the
states between 2 and $M+1$. In the OFF state, instantaneous
transmission rate exceeds the instantaneous channel capacity and
reliable communication can not be achieved. Therefore, the service
rate in this state is effectively zero.

In the next result, we provide the effective capacity for the
cognitive radio channel and state transition model described in the
previous section.

\begin{theo}\label{theo:effectcap}
For the cognitive radio channel with the state transition model
given in Section \ref{subsec:state}, the normalized effective
capacity (in bits/s/Hz) under the average interference power constraint (\ref{average interference power}) is given by
\begin{align}\label{eq:effectivecap}
\hspace{-.8cm}R_{E}(\tSNR,\theta)=-\frac{1}{\theta
TB_{c}}\hspace{-.1cm}\max_{\substack{\alpha^{M-1}\rho P_{d}
E_{}\left\{P_{1}z_{sp}\right\}\\+\sum_{k=1}^{M}\alpha^{M-k}\left(1-\alpha\right)^{k-1}\rho(1-P_{d})\frac{M!}{(M-k)!k!}
E_{k}\left\{P_{2}z_{sp}\right\}\\
\leq I_{avg}}}\hspace{-1.2cm}
\log_{e} \bigg(p_{1}E_{}\left\{e^{-(T-N)\theta
r_{1}}\right\}+\sum_{k=1}^{M}p_{k+1}E_{k}\left\{e^{-(T-N)\theta
r_{2}}\right\} + p_{M+2}\bigg).
\end{align}
Above, $p_{k}$ for $k=1,2,\dots,M+2$ denote the state transition probabilities defined in (\ref{eq:transitionprob1}), (\ref{eq:transitionprobk}), and (\ref{eq:transitionprobM+2}) in Section \ref{subsec:state}. Note also that the maximization is with respect to the power adaptation policies $P_1$ and $P_2$.
\end{theo}

\emph{Remark:} In the effective capacity expression (\ref{eq:effectivecap}), the expectation $E_{}\left\{P_{1}z_{sp}\right\}$ in the constraint and $E_{}\left\{e^{-(T-N)\theta r_{1}}\right\}$ are with respect to the joint distribution of $(z,z_{sp})$ of the channel selected for transmission when all channels are detected busy. The expectations $E_{k}\left\{P_{2}z_{sp}\right\}$ and $E_{k}\left\{e^{-(T-N)\theta
r_{2}}\right\}$ are with respect to the joint distribution of $(z,z_{sp})$ of the channel selected for transmission when $k$ channels are detected as idle.

\emph{Proof of Theorem \ref{theo:effectcap}:} In \cite[Chap. 7, Example 7.2.7]{Performance}, it is
shown for Markov modulated processes that
\begin{gather} \label{eq:theta-envelope}
\frac{\Lambda(\theta)}{\theta} = \frac{1}{\theta} \log_{e}
sp(\phi(\theta)R)
\end{gather}
where $sp(\phi(\theta)R)$ is the spectral radius (i.e., the maximum
of the absolute values of the eigenvalues) of the matrix
$\phi(\theta)R$, $R$ is the transition matrix of the underlying
Markov process, and $\phi(\theta) = \text{diag}(\phi_{1}(\theta),
\ldots, \phi_{M+2}(\theta))$ is a diagonal matrix whose components
are the moment generating functions of the processes in given
states. The rates supported by the cognitive radio channel with the
state transition model described in the previous section can be seen
as a Markov modulated process and hence the setup considered in
\cite{Performance} can be immediately applied to our setting. Since the processes in the states are time-varying transmission rates, we can easily find that $\phi(\theta) =
\text{diag}\left\{E_{}\left\{e^{(T-N)\theta
r_{1}}\right\},E_{1}\left\{e^{(T-N)\theta
r_{2}}\right\},\dots,E_{M}\left\{e^{(T-N)\theta
r_{2}}\right\},1\right\}$. Then, we have
\begin{center}
$\phi(\theta)R=\left[
\begin{array}{cccc}
\phi_{1}(\theta)p_{1} & . & . & \phi_{1}(\theta)p_{M+2} \\
. & \quad & \quad & . \\
. & \quad & \quad & . \\
\phi_{M+2}(\theta)p_{1} & . & . & \phi_{M+2}(\theta)p_{M+2} \\
\end{array}
\right]$.
\end{center}
Since $\phi(\theta)R$ is a matrix with unit rank, we can readily
find that
\begin{align}
sp(\phi(\theta)R)&=\text{trace}\Big(\phi(\theta)R\Big) = \phi_{1}(\theta)p_{1}+\phi_{2}(\theta)p_{2}+\dots+\phi_{M+1}(\theta)p_{M+1}+\phi_{M+2}(\theta)p_{M+2}
\\
&=p_{1}E_{}\left\{e^{(T-N)\theta
r_{1}}\right\}+p_{2}E_{1}\left\{e^{(T-N)\theta
r_{2}}\right\}+\dots+p_{M+1}E_{M}\left\{e^{(T-N)\theta
r_{2}}\right\}+p_{M+2}. \label{eq:sp}
\end{align}
Then, combining (\ref{eq:sp}) with (\ref{eq:theta-envelope}) and
(\ref{exponent}), normalizing the expression with $TB_c$ in order to have the effective capacity in the units of bits/s/Hz, and considering the maximization over power adaptation policies, we reach to the effective capacity formula
given in (\ref{eq:effectivecap}). \hfill $\square$

We would like to note that the effective capacity expression in
(\ref{eq:effectivecap}) is obtained for a given sensing duration
$N$, detection threshold $\gamma$, and QoS exponent $\theta$. In the
next section, we investigate the impact of these parameters on the
effective capacity through numerical analysis. Before the numerical analysis, we first identify below the optimal power adaptation  policies
that the secondary users should employ.

\begin{theo}
The optimal power adaptations for the secondary users under
the constraint given in (\ref{average interference power}) are
\begin{equation}\label{optimalpolicy1}
P_{1}=\left\{
            \begin{array}{ll}
                \frac{\mu_{1}}{z}\left[\left(\frac{z}{z_{sp}\beta_{1}\lambda}\right)^{\frac{1}{c+1}}-1\right],& \hbox{$\frac{z}{z_{sp}}\geq\beta_{1}\lambda$} \\
                0,& \hbox{otherwise}
            \end{array}
        \right.,
\end{equation}
and
\begin{equation}\label{optimalpolicy2}
P_{2}=\left\{
            \begin{array}{ll}
                \frac{\mu_{2}}{z}\left[\left(\frac{z}{z_{sp}\beta_{2}\lambda}\right)^{\frac{1}{c+1}}-1\right],& \hbox{$\frac{z}{z_{sp}}\geq\beta_{2}\lambda$} \\
                0,& \hbox{otherwise}
            \end{array}
        \right.,
\end{equation}
where $\mu_{1}=B_{c}(\sigma_{n}^{2}+\sigma_{s_{p}}^{2})$, $\mu_{2}=\sigma_{n}^{2}B_{c}$,
$c=B_{c}(T-N)\theta/\log_{e}2$, $\beta_{1}=\frac{\mu_{1}\rho
P_{d}}{c\alpha}$ and
$\beta_{2}=\frac{\rho(1-P_{d})\mu_{2}}{c(1-\rho)(1-P_{f})}$.
$\lambda$ is a parameter whose value
can be found numerically by satisfying the constraint (\ref{average interference power}) with equality.
\end{theo}

\emph{Proof:} Since logarithm is a monotonic function, the optimal
power adaptation policies can also be obtained from the following
minimization problem:
\begin{equation}\label{objectivefunc}
\min_{\substack{\alpha^{M-1}\rho P_{d}
E_{}\left\{P_{1}z_{sp}\right\}\\+\sum_{k=1}^{M}\alpha^{M-k}\left(1-\alpha\right)^{k-1}\rho(1-P_{d})\frac{M!}{(M-k)!k!}
E_{k}\left\{P_{2}z_{sp}\right\}\\
\leq I_{avg}}}p_{1}E_{}\left\{e^{-(T-N)\theta
r_{1}}\right\}+\sum_{k=1}^{M}p_{k+1}E_{k}\left\{e^{-(T-N)\theta
r_{2}}\right\}
\end{equation}
It is clear that the objective function in (\ref{objectivefunc}) is
strictly convex and the constraint function in (\ref{average
interference power}) is linear with respect to $P_{1}$ and $P_{2}$ \footnote{Strict convexity follows from the strict concavity of $r_1$ and $r_2$ in (\ref{rate1}) and (\ref{rate2}) with respect to $P_1$ and $P_2$ respectively, strict convexity of the exponential function, and the fact that the nonnegative weighted sum of strictly convex functions is strictly convex \cite[Section 3.2.1]{boyd}.}.
Then, forming the Lagrangian function and setting the derivatives of
the Lagrangian with respect to $P_{1}$ and $P_{2}$ equal to zero, we
obtain:
\begin{align}
&\hspace{-.5cm}\left[\frac{\lambda\rho P_{d}z_{sp}}{\alpha}-\frac{cz}{\mu_{1}}\left(1+\frac{zP_{1}}{\mu_{1}}\right)^{-c-1}\right]\alpha^{M}f_{}(z,z_{sp})=0\label{lamda1}\\
&\hspace{-.5cm}\left[\lambda\rho(1-P_{d})z_{sp}-\frac{c(1-\rho)(1-P_{f})z}{\mu_{2}}\left(1+\frac{zP_{2}}{\mu_{2}}\right)^{-c-1}\right]\sum_{k=1}^{M}\alpha^{M-k}(1-\alpha)^{k-1}\frac{M!}{(M-k)!k!}f_{k}(z,z_{sp})=0\label{lamda2}
\end{align}
where $\lambda$ is the Lagrange multiplier. Above, $f_{}(z,z_{sp})$ denotes the joint distribution of $(z,z_{sp})$ of the channel selected for transmission when all channels are detected busy. Hence, in this case, the transmission channel is chosen among $M$ channels. Similarly, $f_{k}(z,z_{sp})$  denotes the joint distribution when $k$ channels are detected idle, and the transmission channel is selected out of these $k$ channels. Defining
$\beta_{1}=\frac{\mu_{1}\rho P_{d}}{c\alpha}$ and
$\beta_{2}=\frac{\rho(1-P_{d})\mu_{2}}{c(1-\rho)(1-P_{f})}$, and
solving (\ref{lamda1}) and (\ref{lamda2}), we obtain the optimal
power policies given in (\ref{optimalpolicy1}) and
(\ref{optimalpolicy2}). \hfill $\square$

Now, using the optimal transmission policies given in
(\ref{optimalpolicy1}) and (\ref{optimalpolicy2}), we can express the
effective capacity as follows:
\begin{align}
\hspace{-.7cm}R_{E}(\tSNR,\theta)=&-\frac{1}{\theta TB_{c}}\log_{e}
\Bigg(p_{1}E_{\beta_{1}\lambda}\left\{\left(\frac{z}{z_{sp}\beta_{1}\lambda}\right)^{-\frac{c}{c+1}}\right\}
+\sum_{k=1}^{M}p_{k+1}E_{k,\beta_{2}\lambda}\left\{\left(\frac{z}{z_{sp}\beta_{2}\lambda}\right)^{-\frac{c}{c+1}}\right\}+p_{M+2}\Bigg). \label{eq:optimaleffectivecap}
\end{align}
Above, the subscripts $\beta_{1}\lambda$ and $\beta_{2}\lambda$ in the expectations denote that the lower limits of the integrals are equal these values and not to zero. For instance, $E_{\beta_{1}\lambda}\left\{\left(\frac{z}{z_{sp}\beta_{1}\lambda}\right)^{-\frac{c}{c+1}}\right\} = \int_{\beta_{1}\lambda}^\infty \left(\frac{x}{\beta_{1}\lambda}\right)^{-\frac{c}{c+1}} f_{\frac{z}{z_{sp}}}(x) \, dx$.

Until now, we have not specified the criterion with which the transmission channel is selected from a set of available channels. In (\ref{eq:optimaleffectivecap}), we can easily observe that the effective
capacity depends only on the channel power ratio $\frac{z}{z_{sp}}$, and is increasing with
increasing $\frac{z}{z_{sp}}$ due to the fact that the terms $\left(\frac{z}{z_{sp}\beta_{1}\lambda}\right)^{-\frac{c}{c+1}}$ and $\left(\frac{z}{z_{sp}\beta_{2}\lambda}\right)^{-\frac{c}{c+1}}$ are monotonically decreasing functions of $\frac{z}{z_{sp}}$. Therefore, the criterion for choosing the
transmission band among multiple busy bands unless there is no idle
band detected, or among multiple idle bands if there are idle bands
detected should be based on this ratio of the channel gains. Clearly, the strategy that maximizes the effective capacity is to choose the channel (or equivalently the frequency band) with the highest ratio of
$\frac{z}{z_{sp}}$. This is also intuitively appealing as we want to maximize $z$ to improve the secondary transmission and at the same time minimize $z_{sp}$ to diminish the interference caused to the primary users. Maximizing $\frac{z}{z_{sp}}$ provides us the right balance in the channel selection.

We define $x = \max_{i \in \{1,2,\ldots,M\}}\frac{z_i}{z_{sp,i}}$ where $\frac{z_i}{z_{sp,i}}$ is the ratio of the gains in the $i^{th}$ channel. Assuming that these ratios are independent and identically distributed in different channels, we can express the pdf of $x$ as
\begin{equation}\label{x1}
f_{x}(x)=Mf_{\frac{z}{z_{sp}}}(x)\left[F_{\frac{z}{z_{sp}}}(x)\right]^{M-1},
\end{equation}
where $f_{\frac{z}{z_{sp}}}$ and $F_{\frac{z}{z_{sp}}}$ are the pdf and cumulative distribution function (cdf), respectively, of $\frac{z}{z_{sp}}$, the gain ratio in one channel. Now, the expectation $E_{\beta_{1}\lambda}\left\{\left(\frac{z}{z_{sp}\beta_{1}\lambda}\right)^{-\frac{c}{c+1}}\right\}$, which arises under the assumption that all channels are detected busy and the transmission channel is selected among these $M$ channels, can be evaluated with respect to the distribution in (\ref{x1}).

Similarly, we define $x_k = \max_{i \in \{1,2,\ldots,k\}}\frac{z_i}{z_{sp,i}}$ for $k = 1,\ldots,M$.
 The pdf of $x_{k}$ can be expressed as follows:
\begin{equation}\label{x2}
f_{x_{k}}(x)=kf_{\frac{z}{z_{sp}}}(x)\left[F_{\frac{z}{z_{sp}}}(x)\right]^{k-1}\quad
k=1,2,\dots,M.
\end{equation}
The expectation $E_{k,\beta_{2}\lambda}\left\{\left(\frac{z}{z_{sp}\beta_{2}\lambda}\right)^{-\frac{c}{c+1}}\right\}$ can be evaluated using the distribution in (\ref{x2}).
Finally, after some calculations, we can write the effective capacity
in integral form as
\begin{align}\label{neweffectivecapacity}
R_{E}\left(SNR,\theta\right)&=-\frac{1}{\theta
TB_{c}}\log_{e}\bigg\{M\alpha^{M}\int_{\beta_{1}\lambda}^{\infty}f_{\frac{z}{z_{sp}}}(x)\left[F_{\frac{z}{z_{sp}}}(x)\right]^{M-1}\left[\frac{\beta_{1}\lambda}{x}\right]^\frac{c}{c+1}dx\nonumber\\&(1-\rho)(1-P_{f})M\int_{\beta_{2}\lambda}^{\infty}f_{\frac{z}{z_{sp}}}(x)\left[\alpha+(1-\alpha)F_{\frac{z}{z_{sp}}}(x)\right]^{M-1}\left[\frac{\beta_{2}\lambda}{x}\right]^{\frac{c}{c+1}}dx+p_{M+2}\bigg\}.
\end{align}

\section{Numerical Results}\label{NumericalResults}

In this section, we present numerical results for the effective capacity as a function of the channel sensing reliability (i.e., detection and false alarm probabilities) and the average interference constraints. Throughout the numerical results, we assume that QoS parameter is $\theta = 0.1$, block duration is $T = 1$s, channel sensing duration is $N = 0.1$s, and the prior probability of each channel being busy is $\rho = 0.1$.

Before the numerical analysis, we first provide expressions for the probabilities of operating in each one of the four scenarios described in Section \ref{subsec:state}.
These probabilities are also important metrics
in analyzing the performance. We have
\begin{align}\label{scenarios}
\begin{split}
P\{\text{secondary system is in scenario 1}\} = P_{S_{1}}&=\alpha^{M-1}\rho P_{d},\\
P\{\text{secondary system is in scenario 2}\} = P_{S_{2}}&=\alpha^{M-1}(1-\rho)P_{f},\\
P\{\text{secondary system is in scenario 3}\} = P_{S_{3}} &= \underbrace{\sum_{k=1}^M \left(
\begin{array}{c}
M \\ k
\end{array} \right) \alpha^{M-k} (1-\alpha)^k}_{\substack{\text{probability that at least one channel}\\ \text{is detected as idle}}} \,\,\, \underbrace{\frac{\rho(1-P_{d})}{1-\alpha}}_{\substack{\text{probability that the channel chosen} \\ \text{for transmission is actually busy}\\\text{given that it is detected as idle}}}
\\
\hspace{2cm}&=\frac{(1-\alpha^{M})\rho(1-P_{d})}{1-\alpha},\\
P\{\text{secondary system is in scenario 4}\} = P_{S_{4}}&=\frac{(1-\alpha^{M})(1-\rho)(1-P_{f})}{1-\alpha}.
\end{split}
\end{align}

\begin{figure}
\begin{center}
\includegraphics[width = \figsize\textwidth]{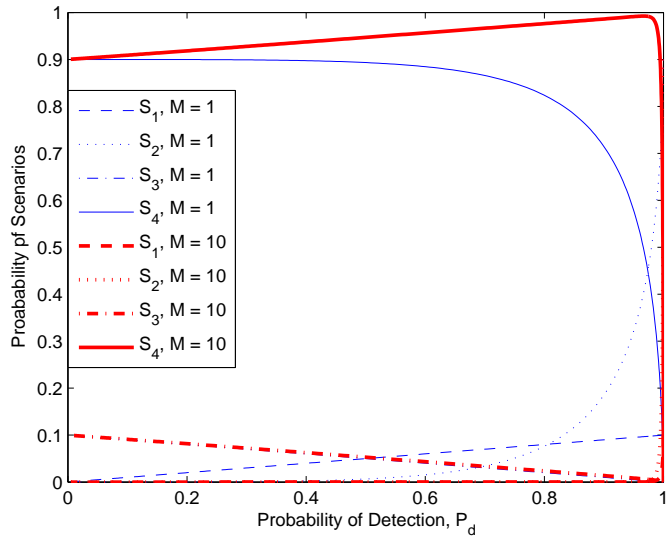}
\caption{Probability of different scenarios vs. probability of detection $P_d$
for different number of channels $M$.}
\label{fig:fig4}
\end{center}
\end{figure}

In Figure \ref{fig:fig4}, we plot these probabilities as a function of the detection probability $P_d$ for two cases in which the number of channels is $M = 1$ and $M = 10$, respectively. As expected, we observe that $P_{S_{1}}$
and $P_{S_{2}}$ decrease with increasing $M$. We also see that $P_{S_{3}}$ and $P_{S_{4}}$ are assuming small values when $P_d$ is very close to 1. Note from Fig. \ref{fig:fig1} that as $P_d$ approaches 1, the false alarm probability $P_f$ increases as well.


\subsection{Rayleigh Fading}
The analysis in the preceding sections apply for arbitrary joint distributions of $z$ and $z_{sp}$ under the mild assumption that the they have finite means (i.e., fading has finite average power). In this subsection, we consider a Rayleigh fading scenario in which the power gains $z$ and
$z_{sp}$ are exponentially distributed. We assume that $z$ and
$z_{sp}$ are mutually independent and each has unit-mean. Then, the pdf and cdf of $\frac{z}{z_{sp}}$ can be
expressed as follows:
\begin{align}\label{rayleigh}
f_{\frac{z}{z_{sp}}}(x)=\frac{1}{(x+1)^{2}}\quad x\geq0 \quad \text{ and } \quad  F_{\frac{z}{z_{sp}}}(x)=\frac{x}{x+1}\quad x\geq0.
\end{align}

\begin{figure}
\begin{center}
\includegraphics[width = \figsize\textwidth]{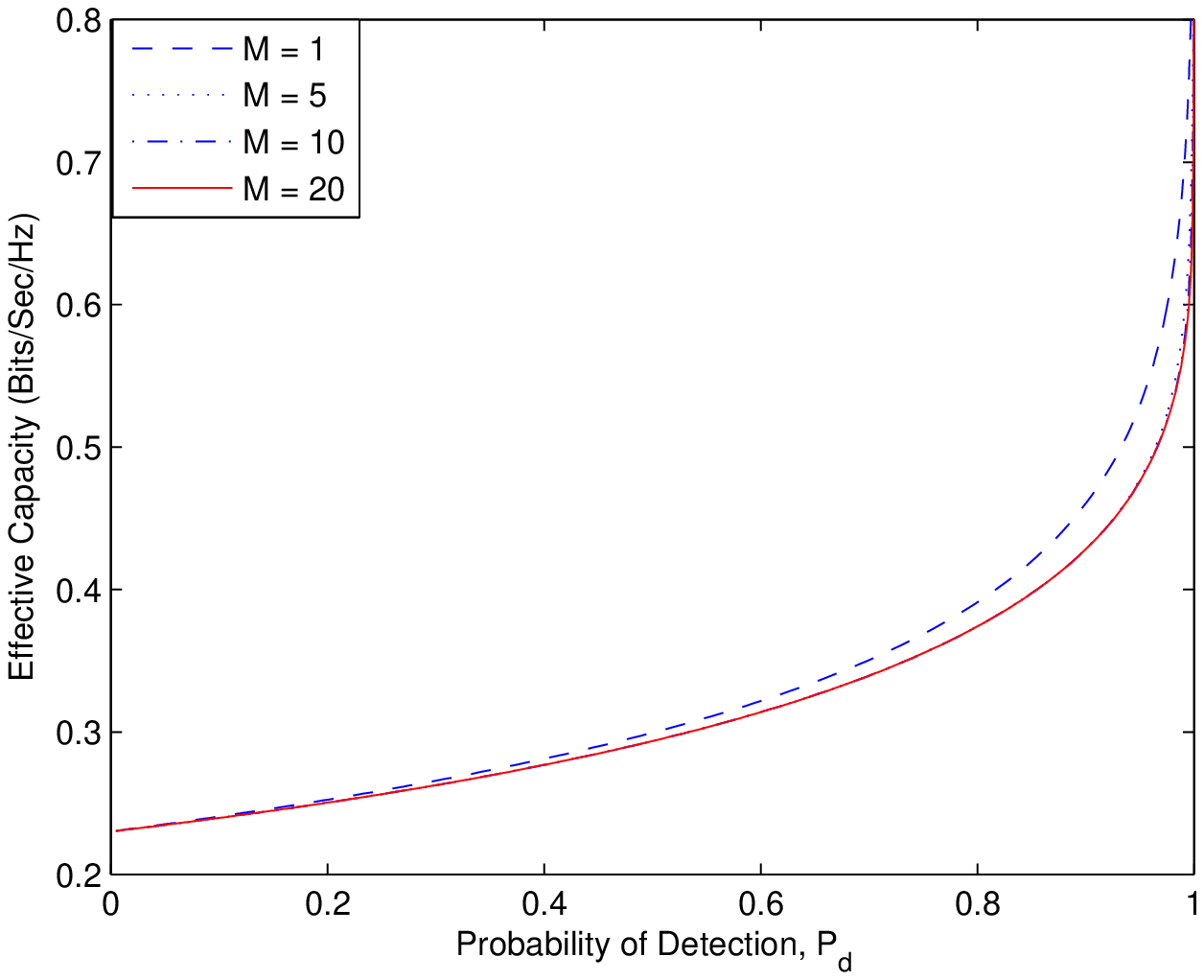}
\caption{Effective capacity vs. probability of detection $P_{d}$
for different number of channels $M$ when $\I_{avg}=0$dB.}
\label{fig:fig5}
\end{center}
\end{figure}

\begin{figure}
\begin{center}
\includegraphics[width = \figsize\textwidth]{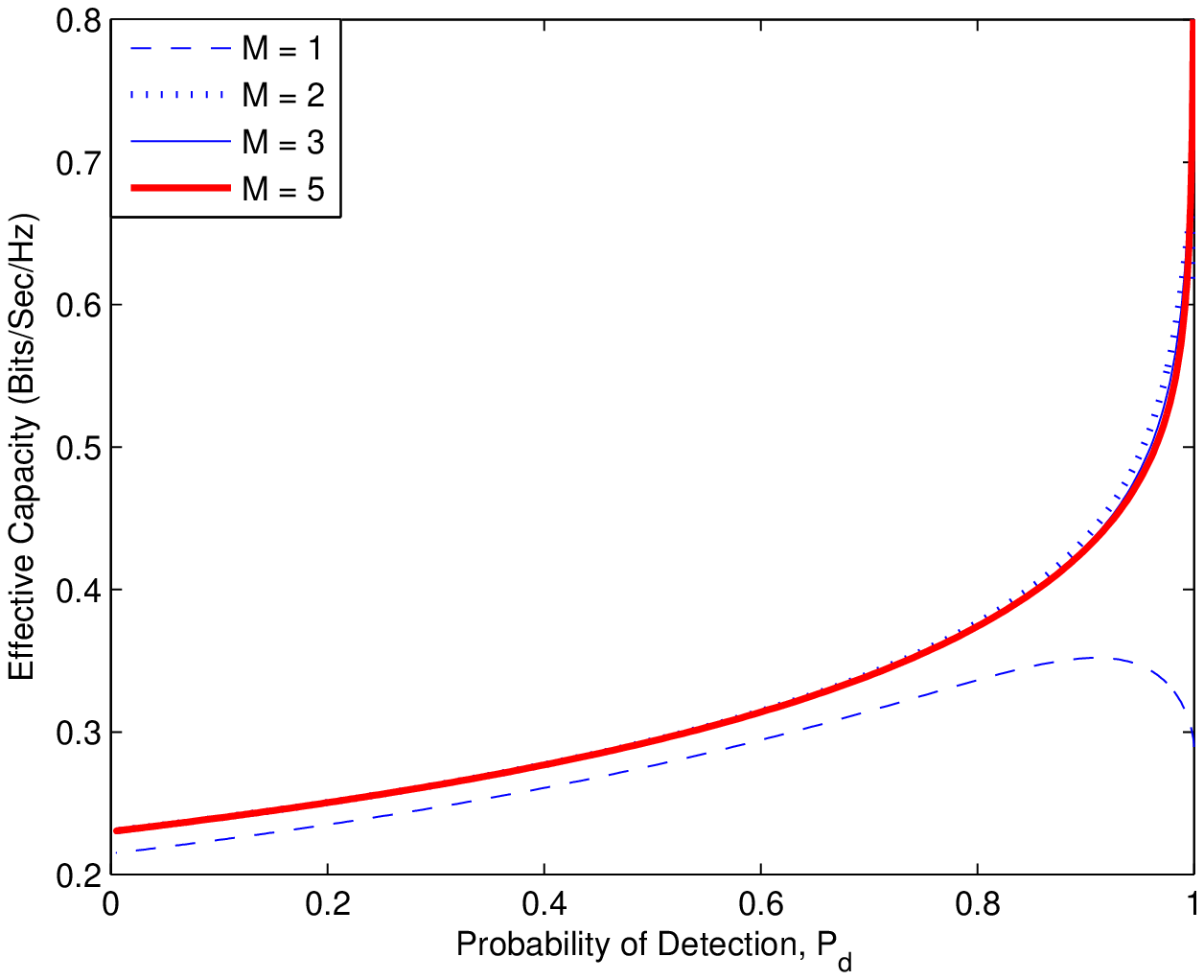}
\caption{Effective capacity vs. probability of detection $P_{d}$
for different number of channels $M$ when $\I_{avg}=-10$dB.}
\label{fig:fig6}
\end{center}
\end{figure}

In Fig. \ref{fig:fig5}, we plot the effective capacity vs. probability of detection,
$P_{d}$, for different number of channels when the average interference power constraint normalized by the noise power is $\I_{avg}(dB) = 10\log_{10} \left(\frac{I_{avg}}{\sigma_{n_p}^{2}B_{c}}\right)=0$dB, where $\sigma_{n_p}^{2}$ is the noise variance at the primary user.
We observe that with increasing $P_{d}$, the effective capacity
is increasing due to the fact more reliable detection of the activity primary users leads to fewer miss-detections and hence the probability of scenario 3 or equivalently the probability of being in state $M+2$, in which the transmission rate is effectively zero, diminishes. We also interestingly see that the highest effective capacity is attained when $M = 1$. Hence, secondary users seem to not benefit from the availability of multiple channels. This is especially pronounced for high values of $P_d$. Although several factors and parameters are in play in determining the value of the effective capacity, one explanation for this observation is that the probabilities of scenarios 1 and 2, in which the secondary users transmit with power $P_1$, decrease with increasing $M$, while the probabilities of scenarios 3 and 4 increase as seen in (\ref{scenarios}). Note that in scenario 3, no reliable communication is possible and transmission rate is effectively zero.  In Fig. \ref{fig:fig6}, we
display similar results when $\I_{avg}=-10$dB. Hence, secondary users operate under more stringent interference constraints. In this case, we note that $M = 2$ gives the highest throughput while the performance with $M = 1$ is strictly suboptimal.

\begin{figure}
\begin{center}
\includegraphics[width = \figsize\textwidth]{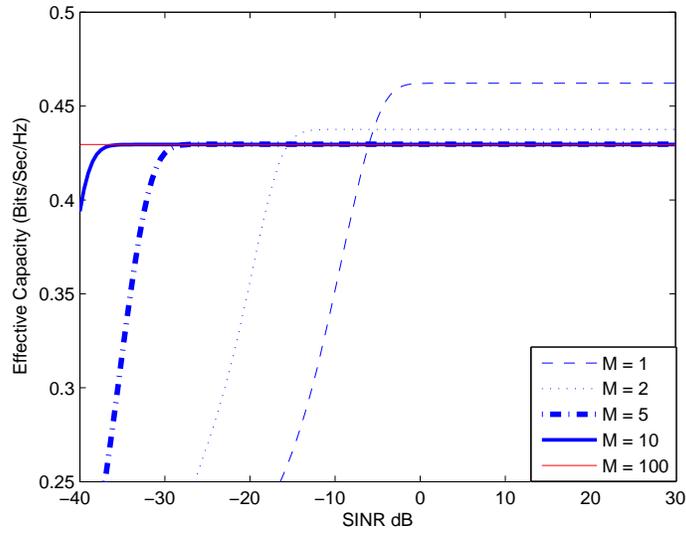}
\caption{Effective capacity vs. $\I_{avg}$ for different values of $M$ when $P_{d}=0.9$ and $P_{f}=0.2$ in the Rayleigh fading channel.}
\label{fig:fig7}
\end{center}
\end{figure}

In
Fig. \ref{fig:fig7}, we show the effective capacities as a
function $\I_{avg}$ (dB) for different values of $M$ when $P_{d}=0.9$ and $P_{f}=0.2$. Confirming our previous observation, we notice that as the interference constraint gets more strict and hence $\I_{avg}$ becomes smaller, a higher value of $M$ is needed to maximize the effective capacity. For instance, $M = 10$ channels are needed when $\I_{avg} < -30$dB. On the other hand, for approximately $\I_{avg} > -6$dB, having $M = 1$ gives the highest throughput.



Above, we have remarked that increasing the number of available channels from which the transmission channel is selected provides no benefit or can even degrade the performance of secondary users under certain conditions. On the other hand, it is important to note that increasing $M$ always brings a benefit to the primary users in the form of decreased probability of interference. In order to quantify this type of gain, we consider below the probability that the channel selected for transmission is actually busy and hence the primary user in this channel experiences interference:
\begin{align}\label{duration}
P_{int}=P\left(\substack{\text{channel selected}\\\text{ for transmission} \\ \text{is actually busy}}\right) &= P\left(\substack{\text{channel selected}\\\text{ for transmission} \\ \text{is actually busy}} \text{ and } \substack{\text{all channels are}\\\text{detected as busy}}\right) + P\left(\substack{\text{channel selected}\\\text{ for transmission} \\ \text{is actually busy}} \text{ and } \substack{\text{at least one channel}\\\text{is detected as idle}}\right)
\\
&= P_{S_1} + P_{S_3}
\\
&=\rho \frac{1-\alpha^{M}-P_{d}+P_{d}\alpha^{M-1}}{1-\alpha}. \label{duration}
\end{align}
Note that $P_{int}$ depends on $P_d$ and also $P_f$ through $\alpha = \rho P_d + (1-\rho)P_f$. It can be easily seen that this interference probability $P_{int}$ decreases with increasing $M$ when $P_d > P_f$.
As $M$ goes to infinity, we have
$\lim_{M\rightarrow \infty} P_{int} = \rho \frac{1-P_{d}}{1-\alpha}.$
Indeed, in this asymptotic regime, $P_{int}$ becomes zero with perfect detection (i.e., with $P_d = 1$). Note that secondary users transmit (if $P_1 > 0$) even when all channels are detected as busy. As $M \to \infty$, the probability of such an event vanishes. Also, having $P_d = 1$ enables the secondary users to avoid scenario 3. Hence, interference is not caused to the primary users.

\begin{figure}
\begin{center}
\includegraphics[width = \figsize\textwidth]{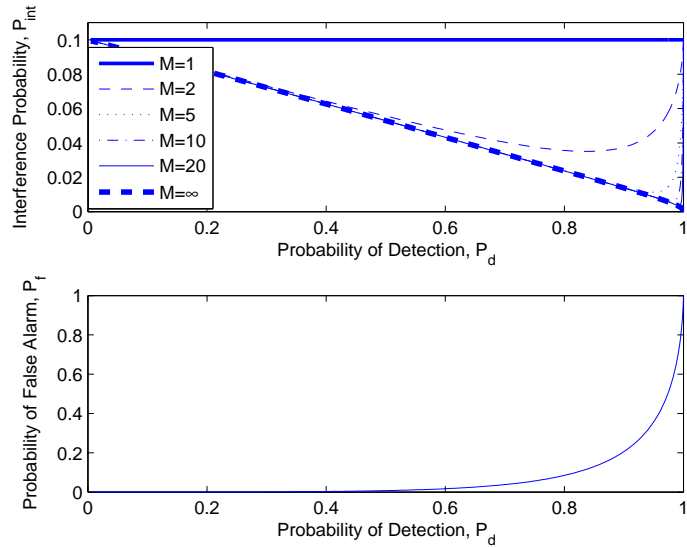}
\caption{$P_{int}$ vs. correct detection probability $P_{d}$ for different number of channels $M$ in the upper figure. False alarm probability $P_f$ vs. correct detection probability $P_d$ in the lower figure.}
\label{fig:fig3}
\end{center}
\end{figure}

In Fig. \ref{fig:fig3}, we plot $P_{int}$ vs. the detection probability for different values of $M$. We also display how the false alarm probability evolves as $P_d$ varies from 0 to 1. It can be easily seen that while $P_{int} = \rho$ when $M = 1$, a smaller $P_{int}$ is achieved for higher values of $M$ unless $P_d = 1$. On the other hand, as also discussed above, we immediately note that $P_{int}$ monotonically decreases to 0 as $P_d$ increases to 1 when $M$ is unbounded (i.e., $M \to \infty$).


\subsection{Nakagami Fading}
\begin{figure}
\begin{center}
\includegraphics[width = \figsize\textwidth]{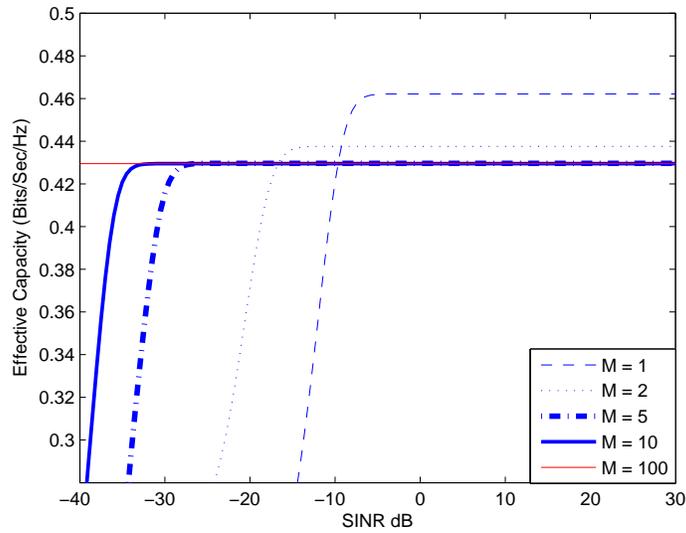}
\caption{Effective capacity vs. $\I_{avg}$ for different values of $M$ when $P_{d}=0.9$ and $P_{f}=0.2$ in the Nakagami-$m$ fading channel with $m = 3$.} \label{fig:fig8}
\end{center}
\end{figure}

Nakagami fading occurs when multipath scattering with
relatively large delay-time spreads occurs. Therefore, Nakagami
distribution matches some empirical data better than many other
distributions do. With this motivation, we also consider Nakagami fading
in our numerical results. The pdf of the Nakagami-$m$
random variable $y=|h|$ is given by
$f_{y}(y)=\frac{2}{\Gamma(m)}\left(\frac{m}{2\sigma_{y}^{2}}\right)^{m}y^{2m-1}e^{-\frac{my^{2}}{2\sigma_{y}^{2}}}$
where $m$ is the number of degrees of freedom. If both $z_{sp}$ and $z$ have the same
number of degrees of freedom, we can express the pdf of $x=\frac{z}{z_{sp}}$ as follows:
\begin{equation}\label{nagakami x}
f_{x}(x)=\frac{\Gamma(2m)x^{m-1}}{(x+1)^{2m}\Gamma(m)^{2}}.
\end{equation}
Note also that Rayleigh fading is a special case of Nakagami fading when
$m=1$. In our experiments, we consider the case in which $m=3$. Now, we can express the cdf of $x$
for $m=3$ as
\begin{equation}\label{Nagamkami 3}
F_{x}(x)=1+\frac{15}{(x+1)^{4}}-\frac{10}{(x+1)^{3}}-\frac{6}{(x+1)^{4}}.
\end{equation}
In Fig. \ref{fig:fig8}, we plot effective capacity vs. $\I_{avg}$ (dB) for different values of $M$ when $P_{d}=0.9$ and $P_{f}=0.2$. Here, we again
observe results similar to those in Fig. \ref{fig:fig7}. We obtain higher throughput by sensing more than one channel in the presence of strict interference constraints on cognitive radios.

\section{Conclusion}\label{Conlusion}

In this paper, we have studied the performance of cognitive transmission under QoS constraints and interference limitations. We have considered a scenario in which secondary users sense multiple channels and then select a single channel for transmission with rate and power that depend on both sensing decisions and fading. We have constructed a state transition model for this cognitive operation. We have meticulously identified possible scenarios and states in which the secondary users operate. These states depend on sensing decisions, true nature of the channels' being busy or idle, and transmission rates being smaller or greater than the instantaneous channel capacity values. We have formulated and imposed an average interference constraint on the secondary users. Under such interference constraints and also statistical QoS limitations in the form of buffer constraints, we have obtained the maximum throughput through the effective capacity formulation. Therefore, we have effectively analyzed the performance in a practically appealing setting in which both the primary and secondary users are provided with certain service guarantees. We have determined the optimal power adaptation strategies and
the optimal channel selection criterion in the sense of maximizing the effective capacity. We have had several interesting observations through our numerical results. We have shown that improving the reliability of channel sensing expectedly increases the throughput. We have noted that sensing multiple channels is beneficial only under relatively strict interference constraints. At the same time, we have remarked that sensing multiple channels can decrease the chances of a primary user being interfered.

%

\end{spacing}

\newpage

\end{document}